\documentclass[epj]{svjour}
\usepackage{graphics}
\usepackage[dvips]{color}
\usepackage{ulem}

\begin{document}
%
\title{The quantum vacuum as the origin of the speed of light}
\author{M. Urban\inst{1}, F. Couchot\inst{1}, X. Sarazin\inst{1} \and A. Djannati-Atai\inst{2}}
%
%
\institute{LAL, Univ Paris-Sud, CNRS/IN2P3, Orsay, France \and APC, Univ Paris Diderot, CNRS/IN2P3, Paris, France}
\date{Received:  / Accepted: }
%
\abstract{
We show that the vacuum permeability $\mu_0$ and permittivity $\epsilon_0$ may originate from the magnetization and the polarization of continuously appearing and disappearing fermion pairs.
We then show that if we simply model the propagation of the photon in vacuum as a series of transient captures within these ephemeral pairs, we can derive a finite photon velocity. Requiring that this velocity is equal to the speed of light constrains our model of vacuum. 
Within this approach, the propagation of a photon is a statistical process at scales much larger than the Planck scale. Therefore we expect its time of flight to fluctuate.
We propose an experimental test of this prediction.
} 

\maketitle
%
%
%

\section{Introduction}
\label{intro}
The vacuum permeability $\mu_0$, the vacuum permittivity  $\epsilon_0$, and the speed of light in vacuum $c$ are widely considered
as being fundamental constants and their values, escaping any physical explanation, are commonly assumed to be invariant in space and time. 
In this paper, we propose a mechanism based upon a "natural" quantum vacuum description which leads to sensible estimations of these three electromagnetic constants.
A consequence of this description is that $\mu_0$, $\epsilon_0$ and $c$ are not fundamental constants but observable parameters of the quantum vacuum:
they can vary if the vacuum properties vary in space or in time.
A similar analysis of the quantum vacuum, as the physical origin of the electromagnetism constants, has been proposed independently by Leuchs, Villar and Sanchez-Soto~\cite{Leuchs}. 
Although the two mechanisms are different, the original idea is the same: the physical electromagnetic constants emerge naturally from the quantum theory.

The paper is organized as follows. First we describe our model of the quantum vacuum filled with continuously appearing and disappearing fermion pairs.
We show how  $\mu_0$ and $\epsilon_0$  originate respectively from the magnetization and the electric polarization of these pairs.
We then derive the photon velocity in vacuum by modeling its propagation as a series of interactions with the pairs. Finally, we predict statistical fluctuations of the transit time of photons across a fixed vacuum path.

%
%

\section{An effective description of the quantum vacuum}
\label{sec:model}

The vacuum is assumed to be filled with continuously appearing and disappearing charged fermion pairs (ephemeral particle-antiparticle pairs).
We consider neither intermediate bosons nor supersymmetric particles. All known species of charged fermions are taken into account:
the three families of charged leptons $e$, $\mu$ and $\tau$ and the three families of quarks ($u$, $d$), ($c$, $s$) and ($t$, $b$),
including their three color states. This gives a total of $21$ pair species, noted $i$.

An ephemeral fermion pair is assumed to be the product of the fusion of two virtual photons of the vacuum. Thus its total electric charge and total
color are null. We suppose also that the spins of the two fermions of a pair are antiparallel, and that they are on their mass shell. The only quantity
which is not conserved is therefore the energy and this is the reason for the limited  lifetime of the pairs. We assume that first order properties can
be deduced assuming that pairs are created with an average energy, not taking into account a full probability density of the pairs kinetic energy.
Likewise, we will neglect the total momentum of the pair.

The average energy  $W_i$ of a pair is taken proportional to its rest mass energy $2 m_i c_{rel}^2$, where $c_{rel}$ is the maximum velocity introduced in the Lorentz transformation. We remind that $c_{rel}$ is not necessarily equal to the speed of light. We note:
\begin{eqnarray}
\label{eq:energy}
W_i\  = K_W \ 2 m_i c_{rel}^2
\end{eqnarray}
where $K_W$ is a constant, assumed to be independent from the fermion type. We take $K_W$ as a free parameter; its value could be calculated if we knew the energy spectrum of the virtual photons together with their probability to create fermion pairs.

As a reminiscence of the Heisenberg principle, the pairs lifetime $\tau_i$ is assumed to be given by
\begin{eqnarray}
\label{eq:tau}
\tau_i = \frac{\hbar}{2W_i} = \frac{1}{K_W}\frac{\hbar}{4 m_i c_{rel}^2}
\end{eqnarray}
We assume that the ephemeral fermion pairs densities $N_i$ are driven by the Pauli Exclusion Principle. Two pairs containing two identical fermions
in the same spin state cannot show up at the same time at the same place. However at a given location we may find 21 charged fermion pairs since
different fermions can superpose spatially. In solid state physics the successful determination of Fermi energies\cite{Kittel} implies that one electron
spin state occupies a hyper volume $h^3$. We assume that concerning the Pauli principle, the ephemeral fermions are similar to the real ones.
Noting $\Delta x_i$ the spacing between identical $i$-type fermions and $p_i$ their average momentum, the one dimension hyper volume is
$p_i\Delta x_i$ and dividing by $h$ should give the number of states which we take as one per spin degree of freedom. The relation between
$p_i$ and $\Delta x_i$ reads $p_i\Delta x_i/h = 1$, or $\Delta x_i=2 \pi \hbar / p_i$.

We can express  $\Delta x_i$ as a function of $W_i$ if we suppose the relativity to hold for the ephemeral pairs
\begin{eqnarray}
\label{eq:dx}
\Delta x_i =\frac{2\pi \hbar c_{rel}}{\sqrt{(W_i/2)^2-(m_i c_{rel}^2)^2}} = \frac{\lambda_{C_i}}{\sqrt{K_W^2-1}}
\end{eqnarray}
where $ \lambda_{C_i}$ is the Compton length associated to fermion $i$ and is given by.
\begin{eqnarray}
\label{eq:lcompton}
\lambda_{C_i} = \frac{h}{m_i c_{rel}} 
\end{eqnarray}

The pair density is defined as:
\begin{eqnarray}
\label{eq:density}
N_i \approx \frac{1}{\Delta x_i^3} = \left(\frac{\sqrt{K_W^2-1}}{\lambda_{C_i}}\right)^3
\end{eqnarray}

Each pair can be produced only in the two fermion-antifermion spin combinations up-down and down-up. We define $N_i$ as the density of pairs for a given spin combination.

Finally, we use the notation $Q_i = q_i/e$, where $q_i$ is the $i$-type fermion electric charge and $e$ the modulus of the electron charge.

%
%

\section{The vacuum permeability}
\label{sec:permeability}

When a torus of a material is energized through a winding carrying a current $I$, the resulting magnetic flux density $B$ is expressed as:  
\begin{eqnarray}
\label{eq:mu-1}
B = \mu_0 n I + \mu_0 M .
\end{eqnarray}
where $n$ is the number of turns per unit of length and $nI$ is the magnetic intensity in $A/m$. $M$ is the corresponding magnetization induced
in the material and is the sum of the induced magnetic moments divided by the corresponding volume. 
In an experiment where the current $I$ is kept a constant and where we lower the quantity of matter in the torus, $B$ decreases. As we remove all matter,
$B$ gets to a non zero value: $B = \mu_0 n I$ showing experimentally that the vacuum is paramagnetic with a vacuum permeability $\mu_0 = 4\pi\ 10^{-7} {N/A^2}$.

We propose a physical mechanism to produce the vacuum permeability from the elementary magnetization of the charged fermion pairs under a magnetic stress.
Each charged ephemeral fermion carries a magnetic moment proportional to the Bohr magneton 
\begin{eqnarray}
\label{eq:magneton}
\mu_i = \frac{e \ Q_i \hbar}{2 m_i}  = \frac{e \ Q_i \ c_{rel} \ \lambda_{C_i}}{4\pi} .
\end{eqnarray}

We assume the orbital moment and the spin of the pair to be zero.
Since the fermion and the anti fermion have opposite electric charges, the pair carries twice the magnetic moment of one fermion.

If no external magnetic field is present, the magnetic moments point randomly in any direction resulting in a null global average magnetic moment. 
In the presence of an external magnetic field $B$, the coupling energy of the $i$-type pair to this field is $-2 \mu_i B \cos \theta$, where $\theta$ is
the angle between the magnetic moment and the magnetic field $\vec{B}$. The energy of the pair is modified by this term and the pair lifetime is therefore
a function of the orientation of its magnetic moment with respect to the applied magnetic field:
\begin{eqnarray}
\label{eq:taumag}
\tau_i(\theta)= \frac{\hbar/2}{W_i - 2 \mu_i B \cos \theta} .
\end{eqnarray}

The pairs having their magnetic moment aligned with the field last a bit longer than the anti-aligned pairs. 
The resulting average magnetic moment $\langle \mathcal{M}_i \rangle$ of a pair is therefore different from zero\footnote{As a referee puts it:``this is a kind of averaged Zeeman effect"} and is aligned with the applied field.
Its value is obtained integrating over $\theta$ with a weight proportional to the pairs lifetime:
\begin{eqnarray}
\label{eq:D}
\langle \mathcal{M}_i \rangle = \frac{\int_0^{\pi} 2\mu_i\ \cos\theta\  \tau_i(\theta)\ 2\pi \sin\theta\ d\theta}{\int_0^{\pi} \tau_i(\theta)\ 2\pi \sin\theta\ d\theta} .
\end{eqnarray}

To first order in $B$, one gets: 
\begin{eqnarray}
\label{eq:magnet}
\langle \mathcal{M}_i \rangle \simeq \frac{4\mu_i^2}{3W_i} B .
\end{eqnarray}

The magnetic moment per unit volume produced by the $i$-type fermions is
$M_i  = {2 N_i \langle \mathcal{M}_i \rangle}$, since one takes into account the two spin states per cell.
The contribution $\tilde{\mu}_{0,i}$ of the $i$-type fermions to the vacuum permeability is thus given by $ B=\tilde{\mu}_{0,i}M_i $ or
${1}/{\tilde{\mu}_{0,i}}={M_i}/{B}$.
Each species of fermions increases the induced magnetization and therefore the magnetic moment. By summing over all pair species, one gets the estimation of the vacuum permeability:
\begin{eqnarray}
\label{eq:permeability-1}
\frac{1}{\tilde{\mu}_0}=\sum_{i}{\frac{M_i}{B}} = c_{rel}^2 \frac{e^2}{6 \pi^2} \sum_{i}{\frac{N_i Q_i^2 {\lambda^2_C}_i}{W_i}}
\end{eqnarray}

Using Eq. (\ref{eq:energy}), (\ref{eq:lcompton}) and (\ref{eq:density}) and  and summing over all pair types, one obtains
\begin{eqnarray}
\label{eq:permeability-2}
\tilde{\mu}_0 = \frac{K_W}{(K_W^2-1)^{3/2}} \frac{24\pi^3\hbar}{c_{rel}\,e^2 \sum_{i}{Q_i^2}}
\end{eqnarray}

The sum $\sum_{i}{Q_i^2}$ is taken over all pair types. 
Within a generation, the absolute values of the electric charges are 1, 2/3 and 1/3 in units of the positron charge.
Thus for one generation  the sum writes $(1+3 \times(4/9+1/9))$. The factor 3 is the number of colours.
Hence, for the three families of the standard model
\begin{eqnarray}
\label{eq:sommeq2}
\sum_{i}{Q_i^2} = 8
\end{eqnarray}

One obtains:
\begin{eqnarray}
\label{eq:permeability-3}
\tilde{\mu}_0 = \frac{K_W}{(K_W^2-1)^{3/2}} \frac{3\pi^3 \hbar}{ c_{rel}\,e^2}
\end{eqnarray}

The calculated vacuum permeability $\tilde{\mu}_0$ is equal to the observed value $\mu_0$ when
\begin{eqnarray}
\label{eq:cadoublev}
\frac{K_W}{(K_W^2-1)^{3/2}}=\mu_0\frac{c_{rel}\,e^2}{3\pi^3\hbar} = \frac{4}{3} \frac{\alpha}{\pi^2}
\end{eqnarray}
which is obtained for $K_W \approx 31.9$\ .

Such a $K_W$ value indicates that the typical fermions are produced in relativistic states. This estimation is based upon a static and average
description of vacuum. A more complete view, including probability densities on pair energy and momentum distributions might allow to give a physical meaning to the $K_W$ value. For instance, $e^+e^-$
pairs with a total energy distributed as $dW/W^2$ up to $W_{max}$ would give an apparent $K_W$ of the order of 
$$K_W \simeq {\rm Log} \left( \frac{W_{max}}{2 m_e c^2} \right) \simeq 51$$
if $W_{max}$ corresponds to the Planck energy. 

%
%
\section{The vacuum permittivity}
\label{sec:permittivity}
Consider a parallel-plate capacitor with a gas inside. When the pressure of the gas decreases, the capacitance decreases too until there are no more molecules
in between the plates. The strange thing is that the capacitance is not zero when we hit the vacuum. In fact the capacitance has a very sizeable value as if the vacuum
were a usual material body. The dielectric constant of a medium is coming from the existence of opposite electric charges that can be separated under the influence of
an applied electric field $\vec{E}$. Furthermore the opposite charges separation stays finite because they are bound in a molecule. These opposite translations result
in opposite charges appearing on the dielectric surfaces in regard to the metallic plates. This leads to a decrease of the effective charge, which implies a decrease
of the voltage across the dielectric slab and finally to an increase of the capacitance. In our model of the vacuum the ephemeral charged fermion pairs are the pairs
of opposite charge and the separation stays finite because the electric field acts only during the  lifetime of the pairs. In an absolute empty vacuum, the induced charges
would be null because there would be no charges to be separated and the capacitance of a parallel-plate capacitor would go to zero when one removes all molecules from the gas. 

We show here that our vacuum filled by ephemeral fermions causes its electric charges to be separated and to appear at the level of $5.10^7$ electron charges per square
meter under an electric stress $E = 1\ V/m$. The mechanism is similar to the one proposed for the permeability. However, we must assume here that every fermion-antifermion
ephemeral pair of the $i$-type bears a mean electric dipole $d_i$ given by:
\begin{eqnarray}
\label{eq:elecdipole}
\vec{d_i} = Q_i e \vec{\delta_i} 
\end{eqnarray}
where $\delta_i$ is the average separation between the two fermions of the pair. 
We assume that this separation does not depend upon the fermion momentum and we use the reduced Compton wavelength of the fermion $\lambda_{C_i}/(2\pi)$  as this scale: 
\begin{eqnarray}
\label{eq:deltai}
\delta_i \simeq \frac{\lambda_{C_i}}{2\pi}
\end{eqnarray}

If no external electric field is present, the dipoles point randomly in any direction and their resulting average field is zero. In presence of an external electric field $\vec{E}$,
the mean polarization of these ephemeral fermion pairs produce the observed vacuum permittivity  $\epsilon_0$.
This polarization shows up due to the dipole  lifetime dependence on the electrostatic coupling energy of the dipole to the field. In a field homogeneous at the $\delta_i$ scale,
this energy is $d_i E \cos \theta$ where $\theta$  is the angle between the ephemeral dipole and the electric field $\vec{E}$. The electric field modifies the pairs lifetimes according to their orientation:
\begin{eqnarray}
\label{eq:taudipel}
\tau_i(\theta)= \frac{\hbar/2} {W_i - d_i E \cos \theta}
\end{eqnarray}
As in the magnetostatic case, pairs with a dipole moment aligned with the field last a bit longer than the others. This leads to a non zero
average dipole $\langle D_i \rangle$, which is aligned with the electric field $\vec{E}$ and given, to first order in $E$, by:
\begin{eqnarray}
\label{eq:polar}
\langle D_i \rangle \simeq {{ d_i^2}\over {3W_i}} E
\end{eqnarray}

We estimate the permittivity $\tilde{\epsilon}_{0,i}$ due to $i$-type fermions using the relation $P_i=\tilde{\epsilon}_{0,i}E$,
where the polarization $P_i$ is equal to the dipole density
$P_i=2 N_i \langle D_i \rangle$, since the two spin combinations contribute. Thus:
\begin{eqnarray}
\label{eq:epsi}
\tilde{\epsilon}_{0,i} =2 N_i \frac{\langle D_i \rangle}{E} = 2 N_i e^2 \frac{Q_i^2 \delta_i^2}{3W_i}
\end{eqnarray}

Each species of fermion increases the induced polarization and therefore the vacuum permittivity. By summing over all pair species, one gets the general expression of the vacuum permittivity:
\begin{eqnarray}
\label{eq:epsi0}
\tilde{\epsilon}_{0} = \frac{2 e^2}{3} \sum_{i}{\frac{N_i Q_i^2 \delta_i^2}{W_i}} = \frac{e^2}{6\pi^2}\sum_{i}{\frac{N_i Q_i^2 {\lambda^2_C}_i}{W_i}}
\end{eqnarray}

Expressing the model parameters from Eq.~(\ref{eq:energy}), (\ref{eq:lcompton}), (\ref{eq:density}), and (\ref{eq:sommeq2}), one gets:


\begin{eqnarray}
\label{eq:permittivity}
\tilde{\epsilon}_0 = \frac{(K_W^2-1)^{3/2}}{K_W} \frac{e^2}{3\pi^3 \hbar c_{rel}}
\end{eqnarray}

If we now use the value $K_W$ given in Eq.~(\ref{eq:cadoublev}) obtained from the derivation of the permeabilitty, one gets the right numerical value for the permittivity: $\tilde{\epsilon}_0 = 8.85\, 10^{-12} F/m$.


We verify from Eq.~\ref{eq:permeability-1} and Eq.~\ref{eq:epsi0} that the phase velocity $c_{\phi}$ of an electromagnetic wave in vacuum, given by $c_{\phi}=1/\sqrt{\tilde{\mu}_0  \tilde{\epsilon}_0}$, is equal to $c_{rel}$ the maximum velocity used in special relativity.

We also notice that the permeability and the permittivity do not depend upon the masses of the fermions.
The electric charges and the number of species are the only important parameters. 
This is in opposition to the common idea that the energy density of the vacuum is the dominant factor\cite{Latorre}.


%
%

\section{The propagation of a photon in vacuum}
\label{sec:speedoflight}

We now study the propagation of a real photon in vacuum and we propose a mechanism leading to a finite average photon velocity $\overline{c}_{group}$, which must be equal to $c_{\phi}$ and $c_{rel}$.

When a real photon propagates in vacuum, it interacts with and is temporarily captured by an ephemeral pair. 
As soon as the pair disappears, it releases the photon to its initial energy and momentum state.
The photon continues to propagate with an infinite {\it bare} velocity.
Then the photon interacts again with another ephemeral pair and so on. 
The delay on the photon propagation produced by these successive interactions implies a renormalisation of this {\it bare} velocity to a finite value.

This ``leapfrog" propagation of photons, with instantaneous leaps between pairs, seems natural since
the only length and time scales in vacuum come from fermion pair lifetimes and Compton lengths. This idea is far from being a new one, as can be found for instance in \cite{Dicke}.

By defining $\sigma_i$ as  the cross-section for a real photon to interact and to be trapped by an ephemeral $i$-type pair of fermions, the mean free path of the photon
between two successive such interactions is given by:
\begin{eqnarray}
\label{eq:freepath}
\Lambda_i = \frac{1}{\sigma_i N_i}\ 
\end{eqnarray}
where $N_i$ is the numerical density of virtual $i$-type pairs. 

Travelling a distance $L$ in vacuum leads on average to $\overline{N}_{stop,i}$ interactions on the $i$-type pairs, given by:
\begin{eqnarray}
\label{eq:Nstop}
\overline{N}_{stop,i} = \frac{L}{\Lambda} = L{\sigma_i N_i}
\end{eqnarray}

The photon may encounter the pair any time between its appearence and disappearence. The life time of a pair being $\tau_i$, the photon will be stopped for an average time $\tau_i/2$. 
Each type of fermion pair contributes in increasing the propagation time of the photon. So, the total mean time $\overline{T}$ for a photon to cross a length $L$ is:
\begin{eqnarray}
\label{eq:Tbar}
\overline{T} = \sum_{i}{\overline{N}_{stop,i} \frac{\tau_i}{2}}\ 
\end{eqnarray}

Using Eq. (\ref{eq:Nstop}), we obtain the average photon velocity $\overline{c}_{group}$ as a function of three parameters of the vacuum model: 
\begin{eqnarray}
\label{eq:c-1}
\overline{c}_{group} = \frac{L}{\overline{T} }= \frac{1}{\sum_{i}{\sigma_i N_i \tau_i /2 }}
\end{eqnarray}

Using Eq. (\ref{eq:tau}) and (\ref{eq:density}), we get the expression
\begin{eqnarray}
\label{eq:c-3}
\overline{c}_{group} = \frac{K_W}{\left(K_W^2-1\right)^{3/2}} \ \frac{16 \pi}{\sum_{i}{(\sigma_i/ \lambda^2_{C_i})}} \ c_{rel}
\end{eqnarray}

We now have to define the expression of the cross section $\sigma_i$. We know that it should not depend on the photon energy, otherwise the vacuum would become a dispersive medium. 
Also the interaction of a real photon with a pair must not exchange energy or momentum with the vacuum (for instance, Compton scattering is not possible). 
We assume the cross-section to be proportional to the geometrical cross-section of the pair $\lambda^2_{C_i}$, and to the square of the electric charge $Q_i^2$. 
The cross-section is thus expressed as:
\begin{eqnarray}
\label{eq:sigma}
\sigma_i = k_{\sigma} Q_i^2 \lambda^2_{C_i}
\end{eqnarray}
where $k_{\sigma}$ is a constant which does not depend on the type of fermions.

The calculated photon velocity becomes:
\begin{eqnarray}
\label{eq:c-4}
\overline{c}_{group} = \frac{K_W}{\left(K_W^2-1\right)^{3/2}} \frac{16 \pi}{k_{\sigma} \sum_{i}{Q_i^2}} \ c_{rel}
\end{eqnarray}
Using Eq. (\ref{eq:sommeq2}) and (\ref{eq:cadoublev}), one finally get:
\begin{eqnarray}
\label{eq:c-5}
\overline{c}_{group} = \frac{8 \alpha}{3 \pi k_{\sigma}} \ c_{rel}
\end{eqnarray}

The calculated  velocity $\overline{c}_{group}$ of a photon in vacuum is equal on average to $c_{rel}$ when 
\begin{eqnarray}
\label{eq:ksigma}
k_{\sigma} = \frac{8}{3 \pi} \ \alpha
\end{eqnarray}
It corresponds to a cross-section of $4 \ 10^{-26}$~m$^2$ on an ephemeral electron-positron pair, of the same order as the geometric transversal area of the pair, whose size is given in Eq.~(\ref{eq:deltai}).

We note that the photon velocity depends only on the electrical charge units $Q_i$ of the ephemeral charged fermions present in vacuum.
It depends neither upon their masses, nor upon the vacuum energy density.
We also remark that the average speed of the photon in our medium being $c_{rel}$, the photon propagates, on average, along the light cone.
As such, the effective average speed of the photon is independent of the inertial frame as demanded by relativity. This mechanism relies on the notion of an absolute frame for the vacuum at rest. 
It satisfies special relativity in the Lorentz-Fitzgerald sense. 
This simple model does not preclude some dependence of the speed of light on the photon energy, through trapping cross-section variations.

%
%

\section{Transit time fluctuations}
\label{sec:prediction transit}

An important consequence of our model is that stochastic fluctuations of the propagation time of photons in vacuum are expected, due to the fluctuations
of the number of interactions of the photon with the virtual pairs and to the capture time fluctuations. 

These stochastic fluctuations are not expected in standard Quantum Electrodynamics, which considers $c$ as a given, non fluctuating, quantity.
Quantum gravity theories predict also stochastic fluctuations of the propagation time of photons~\cite{Yu-Ford}~\cite{Ellis}. 
It has been also recently predicted that the non commutative geometry at the Planck scale should produce a spatially coherent space-time jitter\cite{Hogan}.
We show here that our effective model of photon propagation predicts fluctuations at a higher scale, which makes it experimentally testable with femtosecond pulses.

The propagation time $T$ of a photon which crosses a distance $L$ of vacuum is:
\begin{eqnarray}
\label{eq:transit}
T = \sum_{i=1}\sum_{k=1}^{N_{stop,i}} t_{i,k} 
\end{eqnarray}
where $t_{i,k}$ is the duration of the $k^{th}$ interaction on $i$-type pairs and $N_{stop,i}$ the number of such interactions.
The variance of $T$, due to the statistical fluctuations of the number of interactions and the fluctuation of the capture time is given by:
\begin{eqnarray}
\label{eq:sig1}
\sigma_T^2 = \sum_{i} \left( {\sigma_{N_{stop,i}}^2 \ \overline{t}_{stop,i}^2 + \overline{N}_{stop,i} \ \sigma^2_{t,i}} \right)
\end{eqnarray}
where $\overline{t}_{stop,i} = \tau_i/2$ is the average stop time on a $i$-type pair, $\sigma^2_{t,i}= \tau_i^2/12$ its variance, and $\sigma_{N_{stop,i}}^2 = N_{stop,i}$
the variance of the number of interactions. Hence:
\begin{eqnarray}
\label{eq:sig2}
\sigma_T^2 = \frac{1}{3}\sum_{i}{\overline{N}_{stop,i} \tau_i^2} = \frac{L}{3} \sum_{i}{\sigma_i N_i \tau_i^2}
\end{eqnarray}
Once reduced, the current term of the sum is proportional to $\lambda_{C_i}$. Therefore the fluctuations of the propagation time are dominated by virtual $e^+e^-$ pairs.
Neglecting the other fermion species, and using $\sigma_e N_e\tau_e/2=1/(8c)$, one gets
\begin{eqnarray}
\label{eq:formulesigmat2}
\sigma_T^2 = \frac{\tau_e \ L}{12c}= \frac{\lambda_{C_e}L}{96 \pi K_W c^2}
\end{eqnarray}
So
\begin{eqnarray}
\label{eq:fluctuation}
{\sigma_T} = \sqrt{\frac{L}{c}}\sqrt{\frac{\lambda_{C_e}}{c}}\frac{1}{\sqrt{96 \pi K_W}} 
\end{eqnarray}

In our simple model where $K_W=31.9$, the predicted fluctuation is:
\begin{eqnarray}
\label{eq:fluctuation-2}
\sigma_T \approx 5 \ 10^{-2} \ \mathrm{fs.m}^{-1/2}
\end{eqnarray}

We note that the fluctuations vary as the square root of the distance $L$ of vacuum crossed by the photons and are {\it a priori} independent of the energy of the photons. It is in contrast with expected fluctuations calculated in the frame of Quantum-Gravitational Diffusion~\cite{Ellis}, which vary linearly with both the distance $L$ and the energy of the photons.

A way to search for these fluctuations is to measure a possible time broadening of a light pulse travelling a distance $L$ of vacuum. This may be done using
observations of brief astrophysical events, or dedicated laboratory experiments. 

The strongest direct constraint from astrophysical observations is obtained with the very bright GRB 090510, detected by the Fermi Gamma-ray Space Telescope\cite{Abdo},
at MeV and GeV energy scale. 
It presents short spikes in the 8~keV$-$5~MeV energy range, with the narrowest widths of the order of 4~ms (rms).
Observation of the optical after glow, a few days later by ground based spectroscopic telescopes gives a common redshift of $z = 0.9$. 
This corresponds to a distance, using standard cosmological parameters, of about $2\, 10^{26}$~m. 
Assuming that the observed width is correlated to the emission properties, this sets a limit for transit time fluctuations $\sigma_T$ of about 0.3~fs.m$^{-1/2}$.
It is important to notice that there is no expected dispersion of the bursts in the interstellar medium at this energy scale. 
If we move six orders of magnitude down in distances we arrive to kpc and pulsars.
Short microbursts contained in main pulses from the Crab pulsar have been recently observed at the Arecibo Observatory telescope at 5 GHz\cite{Crab-pulsar-2010}.
The frequency-dependent delay caused by dispersive propagation through the interstellar plasma is corrected using a coherent dispersion removal technique. 
The mean time width of these microbursts after dedispersion is about 1~$\mathrm{\mu}$s, much larger than the expected broadening caused by interstellar scattering.
Assuming again that the observed width is correlated to the emission properties, this sets a limit for transit time fluctuations of about 0.2~fs.m$^{-1/2}$.

The very fact that the predicted statistical fluctuations should go like the square root of the distance implies the exciting idea that experiments on Earth do compete
with astrophysical constraints since we expect fluctuations in the femtosecond range at the kilometer scale.
An experimental setup using femtosecond laser pulses sent to a 100~m long multi-pass vacuum cavity equipped with metallic mirrors could be able to detect this phenomenon.  
With appropriate mirrors with no dispersion on the reflections, a pulse with an initial time width of 9~fs (FWHM) would be broadened after 30 round trips in the cavity,
to an output time width of $\sim 13$~fs (FWHM). An accurate autocorrelation measurement could detect this effect.

\section{Conclusions}

We describe the ground state of the unperturbed vacuum as containing a finite density of charged ephemeral fermions antifermions pairs.  
Within this framework, $\epsilon_0$ and $\mu_0$ originate simply from the electric polarization and from the magnetization of these pairs when the vacuum is stressed by an electrostatic or a magnetostatic field respectively. 
Our calculated values for $\epsilon_0$ and $\mu_0$ are equal to the measured values when the fermion pairs are produced with an average energy of about $30$ times their rest mass.
The finite speed of a photon is due to its successive transient captures by these virtual particles.
This model, which proposes a quantum origin to the electromagnetic constants $\epsilon_0$ and $\mu_0$ and to the speed of light, is self consistent: 
the average velocity of the photon $\overline{c}_{group}$, the phase velocity of the electromagnetic wave $c_{\phi}$, given by $c_{\phi}=1/\sqrt{\mu_0 \epsilon_0}$, and the maximum velocity used in special relativity $c_{rel}$ are equal.
The propagation of a photon being a statistical process, we predict fluctuations of its time of flight of the order of $0.05 fs/\sqrt{m}$.
This could be within the grasp of modern experimental techniques and we plan to assemble such an experiment.

\section*{Acknowledgments}

It is a pleasure acknowledging helpful discussions with Gerd Leuchs. 
The authors thank also J.P. Chambaret , I. Cognard, J. Ha\"{\i}ssinski, P. Indelicato, J. Kaplan, C. Rizzo, P. Wolf and F. Zomer for fruitful discussions,
and  N. Bhat, E. Constant, J. Degert, E. Freysz, J. Oberl\' e and M. Tondusson for their collaboration on the experimental aspects.
This work has benefited from a GRAM\footnote{CNRS INSU/INP program with CNES \& ONERA participations (Action Sp\' ecifique "Gravitation, R\'ef\'erences, Astronomie, M\'etrologie")} funding.

%

\end{document}